\documentclass[sigconf]{acmart}
\usepackage{booktabs}
\usepackage{subfig}
\usepackage{url} 
\usepackage{booktabs}
\usepackage{graphicx}
\usepackage{amsmath}
\usepackage{amssymb}
\usepackage{multirow}
\usepackage{url}
\usepackage{tabularx}
\usepackage{color}
\usepackage{enumitem}
\usepackage{footnote}
\makesavenoteenv{tabular}
\makesavenoteenv{table}

\definecolor{amaranth}{rgb}{0.9, 0.17, 0.31}

\def\mh#1{{\color{black} #1}}
\newcommand\tabitem{\makebox[1em][r]{\textbullet~}}

\copyrightyear{2018} 
\acmYear{2018} 
\setcopyright{acmcopyright}
\acmConference[SIGIR '18]{The 41st International ACM SIGIR Conference on Research \& Development in Information Retrieval}{July 8--12, 2018}{Ann Arbor, MI, USA}
\acmBooktitle{SIGIR '18: The 41st International ACM SIGIR Conference on Research \& Development in Information Retrieval, July 8--12, 2018, Ann Arbor, MI, USA}
\acmPrice{15.00}
\acmDOI{10.1145/3209978.3210143}
\acmISBN{978-1-4503-5657-2/18/07}

\title{Explaining Controversy on Social Media via Stance Summarization}

\begin{document}


\author{Myungha Jang}
\email{mhjang@cs.umass.edu}
\affiliation{\institution{
        Center for Intelligent Information Retrieval \\ 
        College of Information and Computer Sciences \\ 
        University of Massachusetts Amherst
    }}

\author{James Allan}
\email{allan@cs.umass.edu}
\affiliation{\institution{
        Center for Intelligent Information Retrieval \\ 
        College of Information and Computer Sciences \\ 
        University of Massachusetts Amherst
    }}

\begin{abstract}
    In an era in which new controversies rapidly emerge and evolve on social media, navigating social media platforms to learn about a new controversy can be an overwhelming task. In this light, there has been significant work that studies how to identify and measure controversy online. However, we currently lack a tool for effectively understanding controversy in social media. For example, users have to manually examine postings to find the arguments of conflicting stances that make up the controversy. 
    
    In this paper, we study methods to generate a stance-aware summary that explains a given controversy by collecting arguments of two conflicting stances. We focus on Twitter and treat stance summarization as a ranking problem of finding the top $k$ tweets that best summarize the two conflicting stances of a controversial topic. We formalize the characteristics of a good stance summary and propose a ranking model accordingly. We first evaluate our methods on five controversial topics on Twitter. Our user evaluation shows that our methods consistently outperform other baseline techniques in generating a summary that explains the given controversy. 
\end{abstract}
\maketitle


\section{Introduction}
Online controversies often emerge and evolve quickly due to the nature of social media. These platforms such as Twitter and Facebook encourage users to be concise and allow them to be casual, requiring less effort to post something compared to other platforms, such as Wikipedia and blogs.  
While existing techniques enable us to identify \textit{whether} a topic is controversial, understanding \textit{why} it is controversial is still left as work for users. For instance, consider the following scenario: A person discovers a new hashtag movement \texttt{\#TakeaKnee}\footnote{This was prevalent during the US national anthem protests that began in 2017.} on Twitter but does not know what it is about or why it is controversial at all. How would she search for people's opinions to better understand the conflicting stances on this topic?

One straightforward approach to this problem would be for the user to search the topic and manually scan the search results until she has read enough conflicting tweets to understand the controversy. However, current search systems make this navigation difficult due to the filter bubble effect. For example, the top posts are likely to be the ones that the user agrees with because her friends liked the posts or she or her friends follow the authors. 

Another strategy for navigating Twitter is to identify a few key hashtags that indicate stances and then search for posts that contain them. As people are forced to write posts under the strict character limit, certain hashtags are utilized as self-created labels for their opinions (e.g., \texttt{\#imwithher} in support of Hillary Clinton or \texttt{\#MAGA} in support of Donald Trump during the 2016 US presidential election). However, because the use of hashtags (even the ones that seemingly contain obvious stances) are known to be noisy \cite{Mohammad2017StanceAS}, the user must still carefully read each tweet. More importantly, she has to go through a large number of noisy tweets that are not useful to understand the controversy while using her own judgment to identify their stance (if they even have one). This process requires substantial effort, critical reasoning, and phenomenal patience. It is clear that users could benefit from automating this process.

We propose a technique that generates a stance-aware summary by selecting the top tweets that best explain a given controversy. Our contributions are as follows:

\begin{itemize}[wide=5pt, leftmargin=\dimexpr\labelwidth + 2\labelsep\relax]
\item{This work appears to be the first unsupervised approach to automatically summarize controversy on social media.}

\item {We characterize what makes a tweet a good summary of controversy, propose three attributes that should be satisfied (i.e., \textit{stance-indicativeness}, \textit{articulation}, and \textit{topic relevance}), and develop methods to estimate them.}

\item {We propose a novel method to estimate the confidence of stance-indication using automatically-obtained stance hashtags, which have typically been used to filter data during manual annotation. }
 
\item {We extensively evaluate various methods including a general summarization technique and our methods via user evaluation and demonstrate that the summaries generated by our methods explain controversy better than the ones by other techniques.}
\end{itemize}

\begin{table*}[h]
\centering
\caption{An example of good (left) and bad (right) summary tweets on ``Abortion'' posted on Nov 4, 2016. The good summaries are selected from our method. Examples of stance hashtags are marked in bold.}
\label{tabl:summary_example}
\begin{tabular}{|l|l|l|}
\cline{1-1} \cline{3-3}
\small
\begin{tabular}[c]{@{}l@{}} 
\tabitem We know it's not okay that for 40 yrs politicians have denied a woman \\
coverage of abortion just because she's poor \textbf{\#\#BeBoldEndHyde }\\ 
\tabitem Read the whole story about \#HarvardSoccer before forming idiotic tweets.\\ 
\tabitem Hillary Clinton voted no to banning late-term abortions, \\
even though over 80\% of  Americans support the ban. \textbf{\#VoteProlife}
\end{tabular} &
& 
\small
\begin{tabular}[c]{@{}l@{}}
\tabitem before I formed you in the womb I knew you jer 1:5\textbf {\#prolife} \\ 
   \textbf{\#Defundpp} [URL] \textbf{\#UnbornLivesMatter} \\ 
\tabitem Abortions: the new fall trend in religious circles {[}URL{]} \\ 
\tabitem Could you imagine crying over ur uni stopping anti abortion protests, \\ 
if you're so pro life then go and f***ing get one?
\end{tabular} \\ 
\cline{1-1} \cline{3-3} 
\end{tabular}
\end{table*}

\section{Related Work} 
This research is related to a few areas: summarization and controversy analysis on social media.

\textbf{Twitter Summarization:} There has been much work on summarizing Twitter postings while most of them focuses on summarizing events \cite{sharif,Duan2012TwitterTS,Chakrabarti2011EventSU,Inouye2011ComparingTS,Yulianti2016TweetbiasedS}. 
Inouye et al. \cite{sumbasic} compare multiple summarization algorithms for Tweet data, and their extensive experiments suggest that the SumBasic algorithm produced the best F1-result in human evaluation, which we also adopt as a summarization baseline in this paper. 
\mh{Some work has focused on generating contrastive summaries from opinionated text \cite{Paul2010SummarizingCV,Guo2015ExpertGuidedCO}. Particularly, Guo et al. studied tweet data to find a controversy summary. They find a pair of contrastive opinions by integrating manually-curated expert opinions and clustering the pairs to generate a summary. However, their model needs curated expert opinions, which requires constant human effort to maintain as the topic evolves. }


\textbf{Controversy Analysis on the Web:} To identify controversial topics in Web documents, some work has demonstrated that identifying relevant Wikipedia pages as well as building a controversy language model is effective \cite{DoriHacohen2015AutomatedCD,Jang2016ImprovingAC,Jang2016ProbabilisticAT}. Several studies then have formally defined a model for controversy detection \cite{Jang2017ModelingCW,formalmodel}. This work defines that controversy should be identified with respect to a given population (or community). Existing work also has focused on identifying controversy on Twitter \cite{Popescu, garimella,likeminded}. Garimella et al. and Fraisier et al. analyze user retweet or follow graphs, which signifies the formation of exclusive communities of like-minded people for controversial topics. Our approach builds on these earlier findings. 
\vspace{-0.1cm}

\section{Approach}
We first discuss what makes a tweet a good summary. We then develop a ranking model that ranks the tweets by how likely a tweet is part of a good summary. Finally, we propose two methods to select the summary from the ranked tweets. 
\vspace{-0.17cm}
\subsection{Ranking Model}
Based on the definition of controversy by previous work, we define a good controversy summary as a description that effectively captures different arguments of two communities that take conflicting stances with each other. After examining many examples (see Table \ref{tabl:summary_example}), we derive three primary components that characterize a good controversy summary tweet. 

\begin{itemize}[wide=5pt, leftmargin=\dimexpr\labelwidth + 2\labelsep\relax]
    \item \textbf{Stance-indicative (S):} A good tweet strongly indicates its stance and is often followed by some particular stance hashtags that are widely used by users from the same stance community. 
    \item \textbf{Articulation (A):} A good tweet is clear, persuasive, and logical. It is also written with proper language. 
    \item \textbf{Topic Relevance (T):} A good tweet is self-explanatory and relevant in the context of a particular topic.
\end{itemize}

For any controversial topic $\mathcal{T}$, we assume that there are always two stances that are in conflict with each other. We denote these stances as $\mathcal{S}_A$ and $\mathcal{S}_B$. Let $\Gamma$ be a summary of a given topic $\mathcal{T}$. We let $\Gamma$ = [$\Gamma_A$, $\Gamma_B$] that denotes the summary of $\mathcal{S}_A$ and $\mathcal{S}_B$, respectively. We define a model that computes whether a tweet $\tau$ is likely to be in the set $\Gamma_A$:
\begin{equation}
    \label{rankingmodel}
    P(\Gamma_A | \tau) = f(P_S(\mathcal{S}_A | \tau), P_A(\tau), P_T(\tau | \mathcal{T}))
\end{equation}
where $P_S(\mathcal{S}_A | \tau)$ computes how likely a tweet indicates $\mathcal{S}_A$, $P_A(\tau)$ computes how articulate the tweet is, and $P_T(\tau | \mathcal{T})$ computes how relevant the tweet is for the topic. 

In the next section, we discuss how to estimate the first two scores. For the topic relevance score, we use the straightforward probability that the tweet sentence was generated from the language model of the given topic, normalized by the tweet length.

\subsection{Estimating Stance-indication}
To estimate stance-indication, we first identify stance hashtags that statistically characterize the stance community. We use the stance hashtags as a proxy to estimate the tweets that indicate the same stance as follows:
\begin{equation*}
    P_S(\mathcal{S}_A | \tau) = \sum_{h \in \mathcal{H}}{P(h | \tau) \cdot P_S(\mathcal{S}_A | h) \cdot P(h)}  
\end{equation*}

Then the score boils down to estimating $P(h|\tau)$, a probability that the tweet includes a given hashtag $h$, and $P_S(\mathcal{S}_A | h)$, a score that indicates how likely $h$ represents $\mathcal{S}_A$. As $\mathcal{S}_A$ and $\mathcal{S}_B$ are mutually exclusive, we penalize ambiguous tweets that are likely to contain stance hashtags of the opposing side by subtracting the score for the opposite stance as follows:
\begin{equation*}
    P_S(\mathcal{S}_A | \tau) = \sum_{h \in \mathcal{H}_A}{\big[P(h | \tau) \cdot P_S(\mathcal{S}_A | h)\big]}  - \sum_{h \in \mathcal{H_B}}{\big[P(h | \tau) \cdot P_S(\mathcal{S}_B | h)\big]}
\end{equation*}
where $\mathcal{H}_A$ and $\mathcal{H}_B$ are the set of stance hashtags that represent $\mathcal{S}_A$ and $\mathcal{S}_B$ respectively. 

\subsubsection{Identifying Stance Hashtags ($\mathcal{H}_A, \mathcal{H}_B$)}
To obtain a set of stance hashtags, we first identify two communities, $C_A$ and $C_B$, each of which represents the group that holds $\mathcal{S}_A$ and $\mathcal{S}_B$, respectively. Following the same procedure introduced by Garimella et al., we construct a user retweet (RT) graph and partition it into two groups \cite{garimella}. We use a simple method that produces only two communities so as not to deal with the extra step of classifying several identified communities to two stances. We leave identifying multiple communities and clustering them into one of the stances of interest to generate the summaries from for the future work. 

Once we identify $C_A$ and $C_B$, we assume that tweets that are written by users from $C_A$ and $C_B$ are likely to indicate $\mathcal{S}_A$ and $\mathcal{S}_B$ respectively. From the two sets of tweets, we compute the information gain \cite{yangfeatureselection} that each hashtag gets for the information of the community class when they are present in the tweets: if we know nothing about the tweet but the hashtag presence, which hashtag best indicates its stance community? Finally, we define $\mathcal{H}_A$, the set of stance hashtag of $\mathcal{S}_A$, as follows:
\begin{equation*}
    \mathcal{H}_A = \{h \in \mathcal{H} | h \in \mbox{\textit{TopN}}(IG, \mathcal{H}) \land \mbox{\textit{freq}}_A(h) > \mbox{\textit{freq}}_B(h)\}
\end{equation*}
where $IG$ is a function that returns the information gain value for the two stance classes for a given hashtag, $\mbox{\textit{freq}}_A$ is the frequency of $h$ in the tweets published from $C_A$, and \mh{${TopN(IG, \mathcal{H})}$ returns the $N$ items that have the highest scores from a given function $IG$ among the items in the given set $\mathcal{H}$}.  
In our experiments, we set $n$ = 30\mh{, which covers a sufficiently high number of tweets in the communiy given that the distribution of hashtag frequency follows the power law \cite{PrezMelin2017ZipfsAB}}. We then let $P_S(S_A | h)$ be the normalized score of $IG(h)$ for all hashtags in the set $\mathcal{H}_A$. 

\subsubsection{Estimating $P(h|\tau)$ via Latent Hashtags}
If we think of hashtags as user-generated annotations, hashtags are incomplete annotations. This means that a lack of a certain hashtag does not necessarily imply that it is not a relevant label. To better utilize hashtags as more accurate signals, we  make hashtags more complete annotations by estimating $P(h | \tau)$ for all hashtags, the probability that tweet $\tau$ generates a hashtag $h$. Therefore, we adopt a character composition model, \textsc{Tweet2Vec}, which finds a vector space representation of tweets to predict user-annotated hashtags \cite{tweet2vec}. The model computes the hashtag posterior probability for a given tweet for all hashtags in their softmax layer in order to find the top hashtag predictions. We use this probability as $P(h | \tau)$ for hashtags that were not explicitly used in the given tweet. 

\subsection{Estimating the level of articulation}
We build a regression model that predicts how well the tweet is written and generate an annotated set of 150 articulate and 150 non-articulate tweets on arbitrary topics. The annotation criteria between the two classes is whether the given tweet is logical, the grammar is sound, and it is written with proper language.  

Similarly, Duan et al. propose a classifier to evaluate the content quality of tweets \cite{Duan2012TwitterTS}. In addition to their features, we include a large set of POS tags that are Twitter-specific provided by TweeboParser \cite{tweeboparser}, N-grams of the POS tags sequence to capture the structural flow of the good sentences, and the ratio of offensive words to penalize usage of inappropriate language, as shown in Table 2. This model is generalizable since the features are not content-specific. We trained a logistic regression model and obtained 89.9\% classification accuracy using 5-fold cross validation.

\begin{table}[h]
\centering
\caption{The features used to train a regression model for predicting the level of tweet articulation.}
\resizebox{\columnwidth}{!}{
\begin{tabular}{@{}ll@{}}
\toprule
\multicolumn{1}{c}{\textbf{Feature}} & \multicolumn{1}{c}{\textbf{Description}}   \\ \midrule
Tweet POS Tags \cite{tweeboparser}                      &                                                    \begin{tabular}[c]{@{}l@{}} 
                                    The ratio of Tweet  POS tags 
                                       \end{tabular}    \\
OOV words \footnote{\url{http://wordlist.aspell.net/12dicts}}                                               & \begin{tabular}[c]{@{}l@{}}
                                    The ratio of words that are not in the dictionary 
                                        \end{tabular}   \\
Offensive Words \footnote{\url{https://www.cs.cmu.edu/~biglou/resources/bad-words.txt}}                  & The ratio of offensive/profane words\\ 
POS Tags N-grams                    & N-grams of Tweet POS Tag sequence \\
Stop words                          & The ratio of stop words    \\
Tweet length                        & The number of characters in a tweet \\ 
Avg. word length                & \begin{tabular}[c]{@{}l@{}} 
                                    The avg. number of characters in tweet words
                                     \end{tabular}    \\ 
\bottomrule
\end{tabular}
}
\end{table}

\subsection{Summary Selection}
We propose two algorithms that aggregate the three probability scores to generate the final $k$ summary, which we set as 10 in our experiments. To produce a final summary to equally cover two stances, both algorithms select $k/2$ tweets from each stance.

\textsc{SumSAT} ranks the tweets by setting the aggregation function $f$ (in Eq.  \ref{rankingmodel}) to be a harmonic mean for the three scores described earlier. \textsc{HashtagSumSAT}, on the other hand, while using the same aggregation function, first identifies the top $k/2$ stance hashtags for each stance and selects the top tweet for each hashtag. While we use a harmonic mean as $f$, any aggregator can be plugged in. The difference of the two algorithms come from whether it globally ranks the tweets or ranks the tweets per each hashtag.

\begin{figure*}[t!]
	\includegraphics[width=0.7\textwidth]{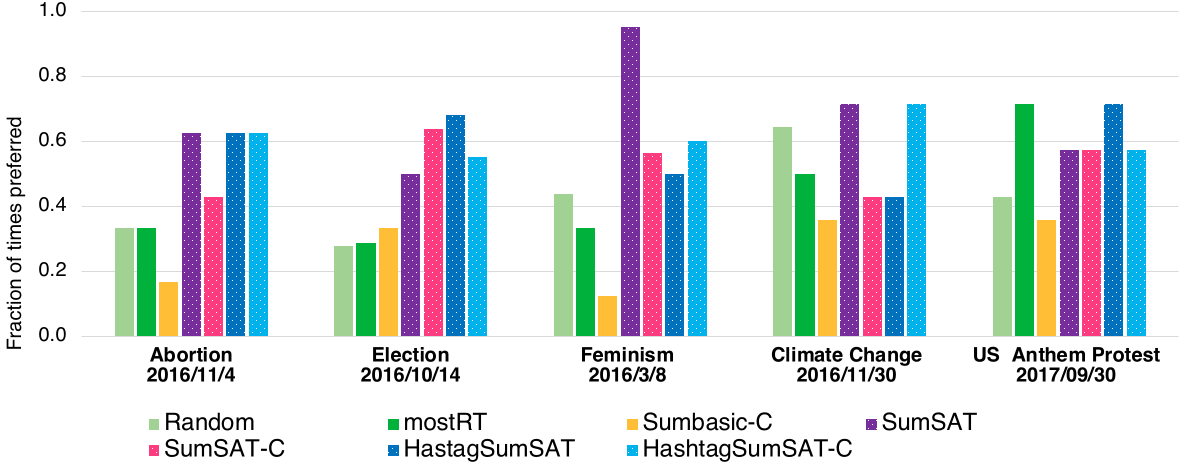}
	\caption{The evaluation results by the topics. The rightmost four bars indicate our methods. We did not include SumBasic in the graph because it was the worst method for all topics, being preferred only 8\% of the time overall.}
    \label{fig:result_bytopics}
\end{figure*}

\section{Evaluation}
We evaluate our methods by running them on real data and conducting user studies to capture the utility of our algorithms.
\vspace{-0.2em}
\subsection{Experiment Setup}
We consider five controversial topics including two short-term, event-based controversies (2016 US Presidential Election and 2017 US National Anthem Protests which we refer to as \texttt{\#TakeAKnee}), and three long-term ethics-related controversies (Abortion, Feminism, and Climate Change). 

Our goal is to generate a summary that can explain why the topic is controversial. For each topic, we generate a pair of summaries and ask 10 participants on Amazon Mechanical Turk which summary better explains the controversy in a double-blind fashion. \mh{A pair of summaries were compared twice by two participants}. The participants could also say that the quality of the two summaries is the same. To observe whether a subset of tweets whose author's stance is identified from the community generates a better quality summary, we experiment with two cases for each algorithm: (1) using all tweets as summary candidates or (2) using only tweets whose author belongs to one of two stance communities we identified. We distinguish the second case by adding `C' (for the community) to the method name. We also generate summaries including the following baseline methods: 

\begin{itemize}[wide=5pt, leftmargin=\dimexpr\labelwidth + 2\labelsep\relax]
    \item \textbf{Random}: A random set of $k$ tweets from a unique set of tweets.
    \item \textbf{MostRT}: The top $k$ most-retweeted tweets in a given day 
    \item \textbf{SumBasic \cite{sumbasic}}: A general summarization technique. We preprocess the tweets to exclude Twitter-specific stop words.  
\end{itemize}

\begin{table}[]
\centering
\caption{The amount of data used to train Tweet2Vec and summary generation. The number in parentheses refers to the number of tweets published by the stance community.}
\label{tabl:dataset}
\resizebox{\columnwidth}{!}{
\begin{tabular}{r|c|c|r|c}
\toprule
\multicolumn{1}{c|}{\multirow{2}{*}{\textbf{Topic}}} & \multicolumn{2}{c|}{\textbf{Tweet2Vec}} & \multicolumn{2}{c}{\textbf{Summary}}                                                                                                 \\ \cline{2-5} 
\multicolumn{1}{c|}{}                                & \textbf{\# Tweets}  & \textbf{\# Users} & \multicolumn{1}{c}{\textbf{\begin{tabular}[c]{@{}c@{}}\# Tweets \\ (\# in C)\end{tabular}}} & \multicolumn{1}{l}{\textbf{RT ratio}} \\ \hline
Election    & 10.8M    & 4.3M & 10000 (4268)  & 70.9\% \\
\#TakeAKnee & 565K     & 692K & 44167 (17217) & 71.1\% \\
Abortion    & 692K     & 539K & 3477 (1262)   & 57.6\% \\
Feminism    & 1.7M     & 1.7M & 50323 (20783) & 41.3\% \\
Climate Change & 546K  & 360K  & 10234 (3915) & 60.1\% \\
\bottomrule
\end{tabular}
}
\end{table}

\subsection{Results and Discussion} 
The evaluation shows that our methods were consistently more effective than other baselines across all five topics (Figure \ref{fig:result_bytopics}). Overall, \textsc{SumSAT} generated the summaries that were preferred the most (68\%) followed by \textsc{HashtagSumSAT-C} (61\%).  


We learned that in identifying and finding stance-indicative tweets, social features are far more important than the content itself. For example, mostRT outperforms a general summarization technique that only considers the text content most of the time. This finding aligns with the findings of the previous study on detecting controversy on Twitter \cite{garimella}. However, depending on the topic and the day, mostRT can also be the worst feature, even worse than random selection as in the case for the topic of Feminism. For example, the top retweets in Feminism include \texttt{`Happy International Women's day!'}. Retweets can often be tweets for entertainment and can easily be dominated by people on one side of the controversy who are more vocal on Twitter.

Our evaluation also suggests that stance hashtags are particularly effective to generate a summary around for event-based controversies, such as the US Election and US Anthem Protest. This is because stance hashtags have been more actively used in these topics as there are usually specific actions that people try to promote or discourage via the hashtags. 

\section{Conclusion}
We introduce and tackle a new task of generating a stance-aware summary to explain controversy on social media. We first characterize three aspects that a desirable summary should satisfy: stance-indication, articulation and topic relevance. We propose a probablistic ranking model that estimates the probability score for each aspect and combines them to find the best summary from the user stance communities. Our human evaluation shows that our summaries are preferred over other baseline summaries in understanding controversy.

\section{Acknowledgements}
\mh{This work was supported in part by the Center for Intelligent Information Retrieval. Any opinions, findings and conclusions or recommendations expressed in this material are those of the authors and do not necessarily reflect those of the sponsor.}

\vspace{-0.1cm}

\bibliographystyle{ACM-Reference-Format}
\bibliography{referencesICTIR} 


\begin{thebibliography}{00}


\ifx \showCODEN    \undefined \def \showCODEN     #1{\unskip}     \fi
\ifx \showDOI      \undefined \def \showDOI       #1{{\tt DOI:}\penalty0{#1}\ }
  \fi
\ifx \showISBNx    \undefined \def \showISBNx     #1{\unskip}     \fi
\ifx \showISBNxiii \undefined \def \showISBNxiii  #1{\unskip}     \fi
\ifx \showISSN     \undefined \def \showISSN      #1{\unskip}     \fi
\ifx \showLCCN     \undefined \def \showLCCN      #1{\unskip}     \fi
\ifx \shownote     \undefined \def \shownote      #1{#1}          \fi
\ifx \showarticletitle \undefined \def \showarticletitle #1{#1}   \fi
\ifx \showURL      \undefined \def \showURL       #1{#1}          \fi
\providecommand\bibfield[2]{#2}
\providecommand\bibinfo[2]{#2}
\providecommand\natexlab[1]{#1}
\providecommand\showeprint[2][]{arXiv:#2}

\bibitem[\protect\citeauthoryear{Chakrabarti and Punera}{Chakrabarti and
  Punera}{2011}]%
        {Chakrabarti2011EventSU}
\bibfield{author}{\bibinfo{person}{Deepayan Chakrabarti} {and}
  \bibinfo{person}{Kunal Punera}.} \bibinfo{year}{2011}\natexlab{}.
\newblock \showarticletitle{Event Summarization Using Tweets} {\em
  (\bibinfo{series}{ICWSM '11})}.
\newblock


\bibitem[\protect\citeauthoryear{Dhingra, Zhou, Fitzpatrick, Muehl, and
  Cohen}{Dhingra et~al\mbox{.}}{2016}]%
        {tweet2vec}
\bibfield{author}{\bibinfo{person}{Bhuwan Dhingra}, \bibinfo{person}{Zhong
  Zhou}, \bibinfo{person}{Dylan Fitzpatrick}, \bibinfo{person}{Michael Muehl},
  {and} \bibinfo{person}{William~W. Cohen}.} \bibinfo{year}{2016}\natexlab{}.
\newblock \showarticletitle{Tweet2Vec: Character-Based Distributed
  Representations for Social Media} {\em (\bibinfo{series}{CoRR '16})}.
\newblock


\bibitem[\protect\citeauthoryear{Dori-Hacohen and Allan}{Dori-Hacohen and
  Allan}{2015}]%
        {DoriHacohen2015AutomatedCD}
\bibfield{author}{\bibinfo{person}{Shiri Dori-Hacohen} {and}
  \bibinfo{person}{James Allan}.} \bibinfo{year}{2015}\natexlab{}.
\newblock \showarticletitle{Automated Controversy Detection on the Web} {\em
  (\bibinfo{series}{ECIR '15})}. \bibinfo{pages}{423--434}.
\newblock


\bibitem[\protect\citeauthoryear{Duan, Chen, Wei, Zhou, and Shum}{Duan
  et~al\mbox{.}}{2012}]%
        {Duan2012TwitterTS}
\bibfield{author}{\bibinfo{person}{Yajuan Duan}, \bibinfo{person}{Zhimin Chen},
  \bibinfo{person}{Furu Wei}, \bibinfo{person}{Ming Zhou}, {and}
  \bibinfo{person}{Harry Shum}.} \bibinfo{year}{2012}\natexlab{}.
\newblock \showarticletitle{Twitter Topic Summarization by Ranking Tweets using
  Social Influence and Content Quality} {\em (\bibinfo{series}{COLING '12})}.
\newblock


\bibitem[\protect\citeauthoryear{Fraisier, Cabanac, Pitarch, Besan\c{c}on, and
  Boughanem}{Fraisier et~al\mbox{.}}{2017}]%
        {likeminded}
\bibfield{author}{\bibinfo{person}{Oph{\'e}lie Fraisier},
  \bibinfo{person}{Guillaume Cabanac}, \bibinfo{person}{Yoann Pitarch},
  \bibinfo{person}{Romaric Besan\c{c}on}, {and} \bibinfo{person}{Mohand
  Boughanem}.} \bibinfo{year}{2017}\natexlab{}.
\newblock \showarticletitle{Uncovering Like-minded Political Communities on
  Twitter} {\em (\bibinfo{series}{ICTIR '17})}. \bibinfo{pages}{261--264}.
\newblock
\showISBNx{978-1-4503-4490-6}


\bibitem[\protect\citeauthoryear{Garimella, De~Francisci~Morales, Gionis, and
  Mathioudakis}{Garimella et~al\mbox{.}}{2016}]%
        {garimella}
\bibfield{author}{\bibinfo{person}{Kiran Garimella}, \bibinfo{person}{Gianmarco
  De~Francisci~Morales}, \bibinfo{person}{Aristides Gionis}, {and}
  \bibinfo{person}{Michael Mathioudakis}.} \bibinfo{year}{2016}\natexlab{}.
\newblock \showarticletitle{Quantifying Controversy in Social Media} {\em
  (\bibinfo{series}{WSDM '10})}. \bibinfo{pages}{33--42}.
\newblock
\showISBNx{978-1-4503-3716-8}


\bibitem[\protect\citeauthoryear{Guo, Lu, Mori, and Blake}{Guo
  et~al\mbox{.}}{2015}]%
        {Guo2015ExpertGuidedCO}
\bibfield{author}{\bibinfo{person}{Jinlong Guo}, \bibinfo{person}{Yujie Lu},
  \bibinfo{person}{Tatsunori Mori}, {and} \bibinfo{person}{Catherine Blake}.}
  \bibinfo{year}{2015}\natexlab{}.
\newblock \showarticletitle{Expert-Guided Contrastive Opinion Summarization for
  Controversial Issues} {\em (\bibinfo{series}{WWW '15})}.
\newblock


\bibitem[\protect\citeauthoryear{Inouye and Kalita}{Inouye and Kalita}{2011}]%
        {Inouye2011ComparingTS}
\bibfield{author}{\bibinfo{person}{David~I. Inouye} {and}
  \bibinfo{person}{Jugal~K. Kalita}.} \bibinfo{year}{2011}\natexlab{}.
\newblock \showarticletitle{Comparing Twitter Summarization Algorithms for
  Multiple Post Summaries}.
\newblock \bibinfo{journal}{{\em PASSAT and SocialCom\/}}
  (\bibinfo{year}{2011}), \bibinfo{pages}{298--306}.
\newblock


\bibitem[\protect\citeauthoryear{Jang and Allan}{Jang and Allan}{2016}]%
        {Jang2016ImprovingAC}
\bibfield{author}{\bibinfo{person}{Myungha Jang} {and} \bibinfo{person}{James
  Allan}.} \bibinfo{year}{2016}\natexlab{}.
\newblock \showarticletitle{Improving Automated Controversy Detection on the
  Web} {\em (\bibinfo{series}{SIGIR '16})}. \bibinfo{pages}{865--868}.
\newblock


\bibitem[\protect\citeauthoryear{Jang, Dori-Hacohen, and Allan}{Jang
  et~al\mbox{.}}{2017}]%
        {Jang2017ModelingCW}
\bibfield{author}{\bibinfo{person}{Myungha Jang}, \bibinfo{person}{Shiri
  Dori-Hacohen}, {and} \bibinfo{person}{James Allan}.}
  \bibinfo{year}{2017}\natexlab{}.
\newblock \showarticletitle{Modeling Controversy within Populations} {\em
  (\bibinfo{series}{ICTIR '17})}. \bibinfo{pages}{141--149}.
\newblock


\bibitem[\protect\citeauthoryear{Jang, Foley, Dori-Hacohen, and Allan}{Jang
  et~al\mbox{.}}{2016}]%
        {Jang2016ProbabilisticAT}
\bibfield{author}{\bibinfo{person}{Myungha Jang}, \bibinfo{person}{John Foley},
  \bibinfo{person}{Shiri Dori-Hacohen}, {and} \bibinfo{person}{James Allan}.}
  \bibinfo{year}{2016}\natexlab{}.
\newblock \showarticletitle{Probabilistic Approaches to Controversy Detection}
  {\em (\bibinfo{series}{CIKM '16})}. \bibinfo{pages}{2069--2072}.
\newblock


\bibitem[\protect\citeauthoryear{Mohammad, Sobhani, and Kiritchenko}{Mohammad
  et~al\mbox{.}}{2017}]%
        {Mohammad2017StanceAS}
\bibfield{author}{\bibinfo{person}{Saif Mohammad}, \bibinfo{person}{Parinaz
  Sobhani}, {and} \bibinfo{person}{Svetlana Kiritchenko}.}
  \bibinfo{year}{2017}\natexlab{}.
\newblock \showarticletitle{Stance and Sentiment in Tweets}.
\newblock \bibinfo{journal}{{\em ACM Trans. Internet Techn.\/}}
  \bibinfo{volume}{17} (\bibinfo{year}{2017}), \bibinfo{pages}{26:1--26:23}.
\newblock


\bibitem[\protect\citeauthoryear{Nenkova and Vanderwende}{Nenkova and
  Vanderwende}{2005}]%
        {sumbasic}
\bibfield{author}{\bibinfo{person}{Ani Nenkova} {and} \bibinfo{person}{Lucy
  Vanderwende}.} \bibinfo{year}{2005}\natexlab{}.
\newblock \bibinfo{booktitle}{{\em The impact of frequency on summarization}}.
\newblock \bibinfo{type}{{T}echnical {R}eport}. \bibinfo{institution}{Microsoft
  Research}.
\newblock


\bibitem[\protect\citeauthoryear{Owoputi, O'Connor, Dyer, Gimpel, Schneider,
  and Smith}{Owoputi et~al\mbox{.}}{2013}]%
        {tweeboparser}
\bibfield{author}{\bibinfo{person}{Olutobi Owoputi}, \bibinfo{person}{Brendan
  O'Connor}, \bibinfo{person}{Chris Dyer}, \bibinfo{person}{Kevin Gimpel},
  \bibinfo{person}{Nathan Schneider}, {and} \bibinfo{person}{Noah~A Smith}.}
  \bibinfo{year}{2013}\natexlab{}.
\newblock \showarticletitle{Improved part-of-speech tagging for online
  conversational text with word clusters} {\em (\bibinfo{series}{HLT-NAACL
  '13})}. \bibinfo{pages}{380--390}.
\newblock


\bibitem[\protect\citeauthoryear{Paul, Zhai, and Girju}{Paul
  et~al\mbox{.}}{2010}]%
        {Paul2010SummarizingCV}
\bibfield{author}{\bibinfo{person}{Michael~J. Paul},
  \bibinfo{person}{ChengXiang Zhai}, {and} \bibinfo{person}{Roxana Girju}.}
  \bibinfo{year}{2010}\natexlab{}.
\newblock \showarticletitle{Summarizing Contrastive Viewpoints in Opinionated
  Text} {\em (\bibinfo{series}{EMNLP '10})}. \bibinfo{pages}{66--76}.
\newblock


\bibitem[\protect\citeauthoryear{P{\'e}rez-Meli{\'a}n, Conejero, and
  Ferri}{P{\'e}rez-Meli{\'a}n et~al\mbox{.}}{2017}]%
        {PrezMelin2017ZipfsAB}
\bibfield{author}{\bibinfo{person}{Jos{\'e}~Alberto P{\'e}rez-Meli{\'a}n},
  \bibinfo{person}{J.~Alberto Conejero}, {and} \bibinfo{person}{C{\'e}sar
  Ferri}.} \bibinfo{year}{2017}\natexlab{}.
\newblock \showarticletitle{Zipf's and Benford's laws in Twitter hashtags} {\em
  (\bibinfo{series}{EACL '17})}. \bibinfo{pages}{84--93}.
\newblock


\bibitem[\protect\citeauthoryear{Popescu and Pennacchiotti}{Popescu and
  Pennacchiotti}{2010}]%
        {Popescu}
\bibfield{author}{\bibinfo{person}{Ana-Maria Popescu} {and}
  \bibinfo{person}{Marco Pennacchiotti}.} \bibinfo{year}{2010}\natexlab{}.
\newblock \showarticletitle{Detecting Controversial Events from Twitter} {\em
  (\bibinfo{series}{CIKM '10})}. \bibinfo{pages}{1873--1876}.
\newblock


\bibitem[\protect\citeauthoryear{Sharifi, Hutton, and Kalita}{Sharifi
  et~al\mbox{.}}{2010}]%
        {sharif}
\bibfield{author}{\bibinfo{person}{Beaux Sharifi},
  \bibinfo{person}{Mark-Anthony Hutton}, {and} \bibinfo{person}{Jugal Kalita}.}
  \bibinfo{year}{2010}\natexlab{}.
\newblock \showarticletitle{Summarizing Microblogs Automatically} {\em
  (\bibinfo{series}{NAACL-HLT '10})}. \bibinfo{address}{Stroudsburg, PA, USA},
  \bibinfo{pages}{685--688}.
\newblock
\showISBNx{1-932432-65-5}


\bibitem[\protect\citeauthoryear{Yang and Pedersen}{Yang and Pedersen}{1997}]%
        {yangfeatureselection}
\bibfield{author}{\bibinfo{person}{Yiming Yang} {and} \bibinfo{person}{Jan~O.
  Pedersen}.} \bibinfo{year}{1997}\natexlab{}.
\newblock \showarticletitle{A Comparative Study on Feature Selection in Text
  Categorization} {\em (\bibinfo{series}{ICML '97})}.
  \bibinfo{pages}{412--420}.
\newblock


\bibitem[\protect\citeauthoryear{Yulianti, Huspi, and Sanderson}{Yulianti
  et~al\mbox{.}}{2016}]%
        {Yulianti2016TweetbiasedS}
\bibfield{author}{\bibinfo{person}{Evi Yulianti}, \bibinfo{person}{Sharin
  Huspi}, {and} \bibinfo{person}{Mark Sanderson}.}
  \bibinfo{year}{2016}\natexlab{}.
\newblock \showarticletitle{Tweet-biased summarization}.
\newblock \bibinfo{journal}{{\em JASIST\/}}  \bibinfo{volume}{67}
  (\bibinfo{year}{2016}), \bibinfo{pages}{1289--1300}.
\newblock


\bibitem[\protect\citeauthoryear{Zielinski, Nielek, Wierzbicki, and
  Jatowt}{Zielinski et~al\mbox{.}}{2018}]%
        {formalmodel}
\bibfield{author}{\bibinfo{person}{Kazimierz Zielinski},
  \bibinfo{person}{Radoslaw Nielek}, \bibinfo{person}{Adam Wierzbicki}, {and}
  \bibinfo{person}{Adam Jatowt}.} \bibinfo{year}{2018}\natexlab{}.
\newblock \showarticletitle{Computing controversy: Formal model and algorithms
  for detecting controversy on Wikipedia and in search queries}.
\newblock \bibinfo{journal}{{\em Information Processing Management\/}}
  \bibinfo{volume}{54}, \bibinfo{number}{1} (\bibinfo{year}{2018}),
  \bibinfo{pages}{14--36}.
\newblock


\end{thebibliography}

\end{document}